\documentclass[preprint]{aastex}

\usepackage{graphicx}
\def\etal{{\it et al.}}
\def\deltam{\Delta M_{\rm e}} 

\def\sm{$M_{\odot}$}

\def\mvt{\hbox{$M_{\rm V}-t$~}}
\def\mv{\hbox{ $M_{\rm V}$}}
\def\mi{\hbox{$M_{\rm i}$}}

\def\mer{ $M_{\rm e}^{\rm R}$~~}

\def\eqmer{ M_{\rm e}^{\rm R}~}

\def\ref{ \noindent\hangindent=1truecm}
\def\yr-1{\hbox{${\rm yr}^{-1}$}}
\def\pc-3{\hbox{${\rm pc}^{-3}$}}

\def\smun{\hbox{$M_\odot$}}

\def\etal{\hbox{${\it\,\, et al.}\,\,$}}

\def\lpl{\hbox{$L_{\rm PL}$}}

\def\mf{\hbox{$M_{\rm f}$}}

\def\lb{\hbox{$L_{\rm B}$}}

\def\lgrav{L^{\pm}_{\rm grav}}

\def\yr{hbox{\rm yr}}
\def\etal{\hbox{\it et al.$\;$}}

\def\tf{\hbox{$t_{\rm F}$}}
\def\ttr{\hbox{$t_{\rm tr}\;$}}

\def\3/2{\hbox{${3\over 2}$}}

\def\tpn{\hbox{$t_{\rm PN}\;$}}

\def\mer{\hbox{$M_{\rm e}^{\rm R}$}}
\def\men{\hbox{$M_{\rm e}^{\rm N}$}}
\def\med{\hbox{$M_{\rm e}^{\rm D}$}}
\def\lsun{\hbox{$L_\odot$}}
\def\lb{\hbox{$L_{\rm B}$}}
\def\lf{\hbox{$L_{\rm F}$}}

\def\teff{\hbox{$T_{\rm eff}$}}

\def\lgrav-{\hbox{$L_{\rm grav}^{-}$}}

\def\smun{\hbox{$M_\odot$}}

\def\yr-1{\hbox{${\rm yr}^{-1}$}}

\def\ha{\relax \ifmmode {\rm H}\alpha\else H$\alpha$\fi}
\def\elc{\relax \ifmmode \epsilon_{\rm Ly c}\else $\epsilon_{\rm Ly c}$\fi}
\def\hb{\relax \ifmmode {\rm H}\beta\else H$\beta$\fi}
\def\alphab{\relax \ifmmode \alpha_{\rm B}\else $\alpha_{\rm B}$\fi}
\def\teff{\relax \ifmmode T_{\rm eff}\else $T_{\rm e}$\fi}
\def\la{\relax \ifmmode L_{\alpha}\else $L_{\alpha}$\fi}
\def\lb{\relax \ifmmode L_{\beta}\else $L_{\beta}$\fi}
\def\betal0{\relax \ifmmode L_{\beta,0}\else $L_{\beta,0}$\fi}
\def\logt{\relax \ifmmode {\rm log}~T_{\rm eff~}\else log~$T_{\rm eff~}$\fi}
\def\logl{\relax \ifmmode {\rm log}~L~\else log~$L~$\fi}
\def\age{\relax \ifmmode {\rm log}~t~\else log~$t~$\fi}
\def\tsw{\relax \ifmmode  t_{\rm sw}\else $t_{\rm sw}$\fi}
\def\mpn{\relax \ifmmode  M_{\rm PN}\else $M_{\rm PN}$\fi~}

\slugcomment{}

\shorttitle{post-AGB Evolution}
\shortauthors{L. Stanghellini \& A. Renzini}

\begin{document}

   \title{Synthetic post-Asymptotic Giant Branch evolution:\\
basic models and applications to disk populations}

\author{Letizia Stanghellini\altaffilmark{1}\affil{Space Telescope Science Institute, 3700 San Martin Drive, Baltimore (MD) 21218; lstanghe@stsci.edu.} 
\and 
Alvio Renzini\altaffilmark{2}
\affil{European Southern Observatory, 2 Karl-Schwarzschild Strasse, 
D-85748 Garching bei M\"unchen (Germany); arenzini@eso.org.}}

\altaffiltext{1}{Affiliated to the Astrophysics Division, Space 
Science Department of ESA. On leave, Bologna Observatory.}
\altaffiltext{2}{{\it on leave}, Bologna University.}

\begin{abstract}
We explore the realm of post-Asymptotic Giant Branch (post-AGB)
stars from a theoretical viewpoint, by 
constructing synthetic population of transition objects, 
proto-Planetary Nebulae, Planetary Nebulae Nuclei, and post-Planetary Nebulae
objects. We use the Montecarlo procedure to filter out the populations 
accordingly to a given set of assumptions. We explore the parameter space 
by studying the effects of the Initial Mass Function (IMF), the
Initial Mass-Final Mass Relation (IMFMR), the transition time (\ttr), 
the envelope mass at the end of the envelope ejection (\mer), the planetary
nebula lifetime \tpn, the hydrogen- and helium-burning phases of the 
central stars.
The results are discussed on the basis of the HR diagram distributions, on the 
\mvt plane, and with mass histograms. 
We found that: (1) the dependence of
the synthetic populations on the assumed IMF and IMFMR is generally 
mild; (2) the
\mer~ indetermination produces very high indeterminations on the \ttr and
thus on the resulting post-AGB populations; (3) the synthetic models 
give a test check for the ratio of He- to H-burning
PNNi.
In this paper, disk post-AGB populations are considered. Future applications will
include Magellanic Clouds PNe, 
and populations of bulges and elliptical galaxies.

\end{abstract}


\section{Introduction}

The post-asymptotic giant branch (post-AGB) phase of evolution, when
stars leave the red giant region to eventually become white dwarf (WD)
remnants, is characterized by a series of intriguing events and
diversities. The beginning of the post-AGB phase is conventionally set at
the
time of cessation of the high mass loss rate episode 
$\dot M \gtrsim 10^{-5} ~\smun \yr-1$ 
which has almost completely stripped the star of
its
hydrogen-rich envelope. After this {\it superwind} (SW) quenching, the
central star shrinks in radius at almost constant luminosity, thus
getting hotter and hotter. Initially it is highly obscured by the
fossil superwind around it, but in $\sim 100-1000$ yr this circumstellar
material becomes optically thin and the star returns to be observable also
at optical wavelengths. The so-called proto-planetary nebula
phase (proto-PN) begins at SW quenching and ceases when the star
becomes hot enough
($\teff\gtrsim 10,000$ K) to
start photo-ionizing the materials that were ejected at relatively low
velocity ($\sim 10-20$ km s$^{-1}$) during the superwind phase. 
As cascade recombination and
free-free emission make the nebular material shining in the optical,
the object has turned into a planetary nebula (PN). The hot star is
now emanating a very fast (several 1000 km s$^{-1}$), radiatively-driven wind
which dynamically interacts with the fossil superwind, occasionally
shaping it in extravagant forms. As nebular expansion continues, the
luminosity and surface brightness of the nebula drop as a result of
its
decreasing emission measure and increasing transparency to the
ionizing photons. A time inevitably comes that its surface brightness
becomes too dim for the nebulosity to be noticed by terrestrial
astronomers,
and the still hot post-AGB star is now slowly evolving towards its final
WD configuration. 

Occasionally, however, before becoming a WD a final
thermal pulse of the still active helium burning shell injects enough
energy
into the outer layers to cause a dramatic expansion of the post-AGB
star, and
sending it back towards the red giant region for a while. Depending on the
details of this metamorphosis, this residual hydrogen-rich
envelope can be engulfed by the underlying convective helium shell, with
its hydrogen being then diluted and burned, and the energy thus
released causing further expansion. A hydrogen-deficient carbon star
- a class of objects including RCrB stars - is thus formed, but even
when the ingestion of the hydrogen envelope does not take place, 
modest wind mass loss suffice to remove virtually all this residual
envelope, thus exposing a bare core with a helium, carbon and oxygen
atmosphere emanating a Wolf-Rayet like spectrum. Therefore, depending
on whether or not post-AGB stars miss to experience such a final thermal
pulse, stars enter the WD stage with or without a hydrogen dominated 
atmosphere, a condition that certainly has to do with the observed
dichotomy of WDs into the DAs and the non-DAs groups (respectively
with and without hydrogen). Life and times of AGB and post-AGB stars have
been extensively reviewed by Iben \& Renzini (1983, IR) and Iben (1993, 1995).

Besides its interest {\it per se}, the post-AGB phase in stellar
evolution has also a number of interesting connections to other
important astrophysical issues. For example
the PN composition provides an essential tool to study the previous 
nucleosynthesis which has taken place during the AGB phase and before,
thus offering important input for modeling the chemical evolution of
galaxies. Moreover, PNs are used as tracers of stellar populations,
e.g., for determining the stellar death rate of galaxies, and are
being used as standard candles in extragalactic distance
determinations. Finally, after neutron stars, post-AGB stars are the
hottest
objects a stellar population can generate, with the most massive of
them coming close to one million K, though for a short while. Their
contribution to the UV and soft X-ray part of the spectrum is
therefore of potential importance, especially in old populations.

The broad scenario sketched above for the post-AGB evolution has now gathered
fairly general consensus - at least for most of its aspects. This
picture
has gradually emerged thanks to an extended  set of stellar
evolutionary calculations, starting with the pioneering
work by Paczy\'nski (1971), and continuing with a series of other relevant
studies (which include: H\"arm \& Schwarzschild 1975; Sch\"onberner
1979, 1983; Iben \etal 1983; Iben 1984, 1985, 1987; Iben \& Tutukov
1984; Iben \& MacDonald 1985, 1986; Wood \& Faulkner 1986; 
Vassiliadis \& Wood 1994; and  Bl\"ocker 1995). 
Still on the theoretical side, several
of the evolutionary transformations mentioned above have been first
sketched in a scattered series of papers (Renzini 1979, 1981a,
1981b, 1982, 1983, 1989, 1990, hereafter P1 through P7), along with a
number of conceptual tools which may be useful for their
understanding. This paper is now
an attempt at presenting these ideas in an orderly and systematic
form. To do so we have proceeded with the construction of
simulated post-AGB populations, as this technique offers an opportunity to
{\it see} in a rather straightforward way what are the observable effects of the
various
theoretical ingredients, and therefore to estimate the relative
uncertainties when - proceeding in the opposite direction - one
attempts to use evolutionary theory to infer astrophysical quantities
from
the observational data.

The paper is organized as follows: In \S 2 we describe our procedure
for the construction of post-AGB synthetic populations, in a way that
makes them easily reproducible. We therein also discuss the role of the
parameters that are at variation in our population synthesis.
In \S 3 we present our results:  
synthetic
post-AGB populations are plotted within the HR diagrams, and $\mv$-age
($\mv$-PN radius) distributions; mass distributions are also shown. 
Our conclusions are summarized in
\S 4, where we also draft future applications. 
In this paper we deal with disk populations with solar
composition. Future work
will extend the parameter space to other populations as well.

\section{Synthetic post-AGB populations: ingredients}

\subsection{Interpolation of the evolutionary tracks}

The stellar models used in this paper 
are interpolation of Vassiliadis and 
Wood (1994, hereafter VW)'s
hydrogen burning evolutionary tracks, calculated for post-AGB masses
between 0.569 and 
0.9 \sm, and solar composition
(from here onward, {\it M} indicates the post-AGB stellar mass). 
To extend the baseline of our population synthesis, we extrapolate to 
lower and higher masses
by using, as templates, the \logt-\logl tracks for $M= 0.546$ and 1.2 \sm~ respectively from Sch\"onberner (1983) and Pacz\'iynski (1971).
We choose the VW database for its homogeneity and wide mass range. 
The VW data set includes also some lower metallicity tracks, which will be used
in a future paper for simulating the Magellanic Cloud post-AGB populations.

In Figure 1 and 2 we plot effective temperature
and
luminosity versus time, as read directly from the VW 
tracks\footnote{In the VW database, all \logt-\age evolutionary tracks start
at t=0,
corresponding to \logt=10,000 K. The curves in Fig. 1 start from the
second evolutionary data point for clarity.}. 
These tracks do not support straightforward interpolation, since
they cross one another in several points.
We divide each 
track into parts roughly representing physical phases, and
then we interpolate the appropriate normalized functions for each phase.

The effective temperature-age curves
have been interpolated as follows.
The six \logt-\age tracks have been divided into four parts, which
hereafter are called (a) {\it H-burning} phase, (b) {\it quenching} phase, 
(c) {\it cooling} phase, and (d) {\it white dwarf} phase. 
In Table 1 we list the characteristics of the VW models. 
Column (1) give the phase, column (2) the mass of the evolutionary models, 
column (3) the characteristic time of each phase and mass.
Note that {\it 
each phase starts at the end of the previous one}, and that the {\it H-burning} phase starts at t=0.
Table 1 also indicates the values of \logt and \logl at the end of each phase, 
in columns (4) and (5). 
From here after, the
capital letters H, Q, C, and W will flag the physical variables at the end 
respectively of the hydrogen burning, the quenching, the cooling, and
the white dwarf phases (e.g., $t_{\rm H}$(0.569 \sm)=$3.215\times10^4$ yr).
For each phase, which correspond to 
a well defined age interval, we have studied the best way to normalize the 
\logt-\age function to eliminate the crossing of the tracks of contiguous
masses.

In Figure 3
we show the temperature evolution on the hydrogen burning
phase.\footnote
{It is understood that all temperatures are in Kelvins, 
all masses in \sm, the stellar luminosities in $L_{\odot}$, and the 
ages in years, unless otherwise noted.}
Here \logt has been plotted against the normalized age

$$t_{\rm nn}=t/t_{\rm H}; \eqno(1)$$ 

the dots represent the normalized evolutionary data points. The interpolation
of the normalized function is straightforward in this phase.
Similarly, Figures 4, 5, and 6 show respectively the \logt curves for the 
quenching, the cooling, and the  
white dwarf phases. In the quenching phase (Fig. 4), it is possible to interpolate 
directly 
\logt versus $t_{\rm nn}$. The normalized age in the quenching phase is:  

$$t_{\rm nn}=(t-t_{\rm H})/(t_{\rm Q}-t_{\rm H}). \eqno(2)$$
 
In the cooling phase (Fig. 5), we use the interpolating function: 
 
$$T_{\rm nn}=\logt/ {\rm log}~ T_{\rm eff,Q} \eqno(3)  $$

with normalized age:

$$t_{\rm nn}=(t-t_{\rm Q})/(t_{\rm C}-t_{\rm Q}). \eqno(4) $$

In the last branch of the evolutionary track, the white dwarf phase (Fig. 6), 
we normalize the effective temperature as:

$$T_{\rm nn}=\logt -{\rm log}~ T_{\rm eff,C}, \eqno(5)  $$

where the normalized age assumes the values of:
 
$$ t_{\rm nn}=(t-t_{\rm C})/(t_{\rm W}-t_{\rm C}) \eqno(6)$$

(note that $T_{\rm nn}$ of Eqs. 3 and 5 do not correspond to an
actual model
temperature).
 
Let us explore the luminosity tracks, for which we use a 
a similar interpolation approach. 
The luminosities against normalized ages are plotted, for phase H, 
in Figure 7. 
The luminosity interpolation of this phase is evidently straightforward.
The quenching phase is very well interpolated by a straight line, calibrated 
for each mass as 

$$L_{\rm nn}=A+ B~[(t-t_{\rm H})/(t_{\rm Q}-t_{\rm H})],
\eqno(7)$$

with $A=0.041$ and $B=1.02$.
The cooling phase normalized luminosity is plotted in Figure 8 for 
tracks corresponding to masses between 0.569 and 0.754 \sm. 
For larger masses, a better interpolation is
achieved by using the power function, with exponent $\alpha$=0.23 for
$0.754 \lesssim M \lesssim 0.8$ \sm, and $\alpha$=0.26 for $M \gtrsim 0.8$ \sm.
Figure 9 shows the white dwarf phase of the luminosity (vs. time) curves, 
where \logl has been plotted against the normalized time,

$$ t_{\rm nn}=(t-t_{\rm C})/(t_{\rm W}-t_{\rm C}). \eqno(8)$$ 

Even in this case, the interpolation is straightforward.

After breaking the \logt and \logl tracks into phases, and
constructing the normalized functions as described above, 
our procedure includes the following steps, for each given
mass and age. (a)
We identify the mass-interval to be used. For example, if $M=0.8$\sm,
we will use the two enclosing tracks for the interpolation, 
corresponding to $0.754$ and $0.9$ \sm.
(b) We identify the phase of the evolution to consider, given the age and
mass. To do so, 
we evaluate $t_{\rm H}$, $t_{\rm Q}$, $t_{\rm C}$, $t_{\rm W}$ for the
given mass. Let us suppose, in our example, that $t=400$ years. We obtain 
respectively 775, 859, and 29,689 yr for the H, Q, and C characteristic times.
This means that at $t=400$ our 0.8 \sm~ star is in the H-burning phase.
(c) We calculate the normalized time, $t_{\rm nn}$. In our example,  
$t_{\rm nn}=0.52$. 
(d) On the appropriate tracks (in our example, Figs. 3 and 7) we read off
\logt and \logl in the appropriate range of $t$ and $M$. In our example,
in Fig. 3 we draw the vertical line $x=0.52$ and read off the \logt values
from the two upper tracks, the four points immediately before and after 
the vertical line. 
(c) We interpolate, on the higher 
and lower mass tracks, and find the \logt and \logl 
that intersect the $x=t_{\rm nn}$ vertical line; (4)
we interpolate {\it vertically} versus log M, where M is the given mass.
 
When evaluated on the masses of the original VW evolutionary tracks, 
the synthetic temperature and luminosity curves are indistinguishable 
from those plotted in Figures 1 and 2.
We have studied the interpolation offsets in detail.
The relative errors in the {\it linear} temperatures,
(T$_{\rm eff,syn}$-T$_{\rm eff,VW}$)/T$_{\rm eff,VW}$ 
\footnote{
The subscripts {\it syn} and {\it VW} refer respectively to the synthetic
and the VW evolutionary tracks.}, 
result to be collimated 
between -0.005 and 0.005 in most evolutionary phases, 
with the exception of the
very fast quenching phase, where the linear errors on the temperature
are rather between -0.02 
and 0.02 for the lower 5 mass tracks, and between -0.02  and 0.1 for
the 0.9 track.
The relative errors in the luminosities, 
(L$_{\rm syn}$-L$_{\rm VW}$)/L$_{\rm VW}$, 
are between -0.01 and 0.01 in most evolutionary phases, 
with the exception of the
quenching phase, where the linear errors on the luminosity
are between -0.15
and 0.15 for the lower 5 mass tracks, and between -0.15 and 0.7 for the 0.9 
track.

When using the interpolated values {\it au lieu} of the evolutionary ones, 
we obtain a perfect substitution. In the quenching phase, 
the substitution would
displace the temperatures by a marginal fraction, and the misplacement of the 
luminosity curves would propagate only in a few percent error on log L.
In all other cases, the substitution reflects in virtually no 
errors.
We conclude that the interpolations are excellent and can be reliably used 
as evolutionary tracks for all stellar masses between 0.569 and 0.9 
\sm. 

We obtain extrapolations to higher and lower masses using the \logt-\logl 
{\it templates} of 0.546 and 1.2 solar masses from Sch\"onberner (1983)
and Pacz\`inski (1971). Only the {\it shape} of the \logt-\logl relation is
used here, not the actual evolution of the physical parameters, which 
are instead extrapolated directly from the 0.596 and 0.9 \sm~ tracks by VW,
for homogeneity.

\subsection{The initial mass function}

The synthetic distribution of post-AGB stars (e.g., on the HR diagram)
depends on the adopted IMF and history of star formation. Note that a
given distribution of post-AGB stars obtained with a certain IMF and star 
formation (SR)
history can be reproduced also with a different IMF provided a
properly tuned SF history is adopted. We have renounced to play with
two
independent parameters - such as the slope of the IMF and the e-folding
time of the SF rate - and have exclusively explored the effect of
changing slope of the {\it actual} initial mass distribution
$\psi(\mi) \propto \mi^{-(1+x)}$. Therefore $x$ is effectively the slope
of the IMF only if a constant star formation rate is assumed. This may
not be a bad approximation for the Galactic Disk (Scalo 1998). 
In our simulations
it is supposed that only stars in the mass range $0.85 \lesssim\mi< 9\smun$
experience the thermal pulses on the AGB (TP-AGB) and the 
post-AGB phases. The upper limit of the mass domain has been chosen in the light of
the carbon ignition limits described in Iben (1995).
We include in our investigation
the classic Salpeter's (1955) IMF, with constant index for each mass range.
Furthermore, we inspect the effects of a IMF with variable exponent, 
such as the 
one by Miller \& Scalo (1979), the low-mass extension by Kroupa \etal
(1991), and the IMF 
recently proposed by Scalo (1998).
In all three cases, the function index changes for masses smaller
or larger than \mi= 1\sm. 

\subsection{The initial mass-final mass relation}

The initial mass-final mass relation (IMFMR) may have a major impact 
on the 
resulting synthetic distributions. 
To illustrate  the case we have constructed simulations adopting
four different IMFMRs, either from theoretical calculations or from
observations: (1) the {\it old} theoretical Renzini \& Voli (1981)
relation as analytically approximated by IR
\footnote{In this paper we flag this IMFMR with `IR'.}; 
(2) a more recent theoretical IMFMR introduced by Ciotti \etal
(1991), which is an attempt at incorporating new important
results of evolutionary calculations; 
(3) the classic empirical IMFMR proposed by Weidemann (1987) on the basis 
of WD masses observed in open clusters;
and (4) a new empirical relation presented by Herwig (1997) that includes 
new observations of cluster white dwarfs in the Pleiades, the Hyades, 
and NGC~3451. 
Figure 10 shows the four IMFMRs (note that in this paper we use $M$=\mf).

A brief justification for the most recent theoretical relation (Ciotti
\etal 1991) is appropriate.
The old theoretical IMFMR (IR) 
was obtained interpolating on an insufficient grid of
stellar models, and assumed a universal core mass-luminosity relation.
Lattanzio (1989) pointed out that the straight line approximation for
the core mass at the first AGB thermal pulse (as a function of $\mi$,
and for $\mi<3\,\smun$)
was much steeper than resulting from actual evolutionary calculations.
Furthermore,
Bl\"ocker \& Sch\"onberner (1991) have found a major breakdown of the core mass-luminosity relation (a key
ingredient in theoretical IMFMRs) for those stellar models in which
the so-called envelope burning process is activated, i.e., for 
 $\mi\gtrsim 3\,\smun$  (cf. Renzini \& Voli
1981). In practice the former effect results in a flattening the IMFMR
{\it below}
$\sim 3\,\smun$, the second one in a flattening of the IMFMR {\it
above} $\sim 3\,\smun$.
The new theoretical IMFMR incorporates
these findings, while keeping the same parameterization of mass loss
processes (wind and superwind) as adopted by Renzini \& Voli, with
$\eta=0.5$ and $b=1$.

On the empirical side, the new relation by Herwig (1997) 
is not very different than
the new theoretical relation. In addition, we should mention that new 
data points by Jeffries (1997) of white dwarfs in NGC~2516 also lie very 
close to Herwig's relation. Furthermore, it should be noted that more recent calculations
presented in conferences (e.g., Lattanzio \& Forestini 1999;
Bl\"ocker 1999) show that the envelope burning occurs for M$>$4 \smun,
bringing the theoretical MIMFR hardly distinguishable from Herwig's (1997).

\subsection{The transition time and the post-AGB envelope mass}

At the TP-AGB phase, the red giant envelope is ejected by the superwind. 
The subsequent evolution of the star to PNN depends in a substantial way
on the amount of envelope mass left on the star at the SW quenching,
\mer. No matter how the transition between the AGB and the PN illumination 
occurs, the remnant envelope mass at the end of the SW plays a central role.
This is just unfortunate when trying to simulate this evolutionary phase:
in fact, \mer~ is not defined by stellar evolution (given the hydrodynamical 
nature of the superwind), and only approximations or guesses can be made 
about its entity and its possible relation to the physical parameters.
\mer~ can indeed be considered a free parameter, in the sense that there is 
no theoretical nor observational constraint to fit it, nor to give it an exact 
dependence on any nebular or stellar parameter.

In principle, the higher the core mass, the higher the stellar luminosity,
and the lower the envelope mass left on the AGB star. But this also depends on 
the thermal pulses that occur at the TP-AGB, and it is hard to know from
first principle how the superwind ejection changes the stellar structure.
We examine later in this section how we choose \mer~ for our simulation.

To understand the transition between the AGB and the PN phases, we
should introduce the timescales involved in the different phases.
Immediately after the superwind quenching, the mass loss due to stellar
wind dominates, until the star detaches itself
from the AGB. The time scale at which this phase occurs is the
wind time scale, and can be written as:

$$t_{\rm w}={\Delta M_{\rm e} \over \dot M}, \eqno(9)$$

where $\deltam$ is the difference between the residual envelope
mass at the superwind quenching, \mer, and the envelope mass at the 
detachment from the AGB (i.e., at a later phase),
\med. 
The wind mass loss rate (MLR) is taken from Reimers (1975), evaluated at the AGB
temperature of 5000 K. 
Reimers' approximation seems reasonable in the considered evolutionary
phase, that is, after the superwind quenching and before the onset of the 
radiation-driven wind (see Bl\"ocker 1995).
In effect, the MLR right after the detachment from the AGB declines as the 
stellar temperature increases (Bl\"ocker 1995); but in the present application,
i.e., to evaluate the transition time, this choice does not affect the results.

For larger temperatures, in the blueward runaway, 
the mass consumption occurs at a nuclear timescale, following the 
Equation:

$$t_{\rm n}={X_{\rm e} \deltam E_{\rm H} \over L}. 
\eqno(10)$$

$\deltam$ is now the difference between the smaller among \mer~ and \med, 
and the amount of envelope mass remaining at the illumination of the nebula, \men; 
the luminosity in the equation is the {\it plateau} luminosity
\footnote{The {\it plateau} luminosity, i.e., the luminosity in the
early post-AGB phase, is $\lpl=56694 \times 
(M-0.5)$ for VW models.};
and the variables $X_{\rm e}$ and $E_{\rm H}$ are the envelope hydrogen
abundance and the energy released by the nuclear burning of one gram 
of hydrogen, respectively.

Finally, the transition time will scale with the shorter among the wind and the nuclear time scales. We set the reciprocal of the transition time to be the 
sum of the reciprocals of the wind and the nuclear times.

The transition
occurs on a thermal timescale in the case in which \mer $\leq$ \men, thus in the
case in which either the nuclear or the wind time is zero. 
The formula to be used then is:

$$t_{\rm th}={G M \eqmer \over L R}, \eqno (11) $$

where L and R are the stellar luminosity and radius (see also P6).

It is worth recalling that the synthetic tracks obtained in $\S$2.1 have their 
zero age points at \logt=4.0, which are approximately reached
when the transition has been 
completed. At 10,000 K, the star is able to ionize hydrogen, although
the complete nebular transparency at the optical wavelengths occurs generally
at slightly higher temperatures (K\"aufl \etal 1993). In all events, we 
use the synthetic tracks from their natural zero point. 

Evolutionary models for post-AGB stars predict the amount of mass available in the stellar
envelope as a function of the effective temperature (Sch\"onberner 1983,
Paczynski 1971). We use Sch\"onberner's models (M$<$~0.65) to obtain 
\med (envelope mass at the detachment) and \men (envelope mass at nebular 
illumination, when T$_{\rm eff}$=10,000 K) as functions of M, by linear interpolation. For M$>$0.65 \sm~ we obtain \med~ and \men~
by scaling 
Paczynski's 1.2 \sm~ model to Sch\"onberner's values. 

As discussed at the beginning of this section, \mer~ is quite unconstrainted,
and its value affords unpredictable variations. We set ourselves to 
explore a wide parameter space with three different assumptions for \mer. 

First, since the evolutionary models have shown that the envelope mass at 
several evolutionary phases in the post-AGB is inversly proportional to 
the core mass,
we could envision the possibility that also at the superwind quenching
\mer~ has an inverse dependence on M. 
We parametrize this case following the usual models 
as guidelines, by taking \mer~ as the envelope mass at the onset of the
post-AGB models.
Figure 11 illustrates how the characteristic timescales vary if
\mer=f(M). The 
transition time peaks for the very small mass models, and then
it declines for larger masses.

Second, we chose a constant \mer, independent of mass. In Figure 12 and 13 we analyze the consequences 
of assuming two values of \mer,
\mer=3$\times 10 ^{-4}$ and \mer=5$\times 10^{-3}$. In the first case (Fig.12),
the transition time follows up close the thermal time scale, then the nuclear
time scale
for m$>$0.85 \sm, to decline for larger masses.
In the second case (Fig. 13), the 
transition time rises to almost 6000 yr for 
M$\sim$0.55 \sm. These two cases with constant \mer~
are illustrated only for the sake of showing extreme cases, but it is
very unrealistic that all stars end the unstable SW phase with exactly
equal \mer.

Third, we use random values of \mer~ for our calculations. In Figure 14 we
show the transition, wind, and nuclear  
time scales for random \mer, with maximum equal to
\mer=$5\times 10^{-3}$.
The transition time peaks at about 0.6 solar masses, then declines.
The same simulation with different maximum \mer~
gives similar results, except for the vertical scale.
Naturally, giving the randomness of the extraction, the population plotted
in Figure 14 is one of the infinite possible extractions with random 
\mer~ lesser than $5\times10^{-3}$ \sm.

\subsection{The final helium-shell flash}

During the post-AGB phase hydrogen is burned in a shell, and therefore
the mass of the helium {\it buffer} zone between the C-O core and the
hydrogen shell keeps increasing. In some cases the increase of the
buffer mass can lead to a last thermal pulse (also called flash) of
the helium burning shell (Sch\"onberner 1979). This happens if 
the star left the AGB with a
sufficiently massive buffer zone, so that its further increase during
the post-AGB allows to reach the critical value
for a last flash to erupt. This chance is therefore related to the {\it
phase} in the AGB thermal pulse cycles at which the SW envelope
ejection takes place. Detailed model calculations show that as a response to
the final flash stars undergo an extended loop in the HR diagram.
The model star would expand back to the AGB very rapidly, then it would evolve 
again as a post-AGB star, powered by helium burning; the duration of
the He-burning phase is about
three
times longer than the fading time of hydrogen-burning post-AGB stars
(Iben 1984). These calculations also show that during the power down
phase
the rate of luminosity decline is nearly constant, as opposed to the
case of
hydrogen burning post-AGB stars in which an abrupt drop follows the
plateau phase.

To incorporate the effect of the final helium-shell flash (FF)
we have proceeded as follows: we assume that
such stars return to $(L,\teff)=(\lpl,10,000$ K) upon a final flash, 
and their luminosity fades then linearly
with time to $L_{\rm F}$ in a time $3\times t_{\rm F}$ (where \tf~ 
is the total fading time for the H-burning star, see $\S$ 2.7.2),
and \lf~ is the luminosity of the corresponding H-burner, 
i.e.:

$$L=\lpl - {\lpl-\lf\over 3\times t_{\rm F}}\cdot t'.\eqno(12)$$

We then locate the star on the \logl - \logt curves 
as if the evolution were three times as slow as H-powered evolution.
This provides a fairly good approximation of the
behavior of the models constructed by Iben (1984).
The fraction of stars experiencing a final flash is not strictly
determined by theory. Iben has calculated the probability that 
a star ignites helium in various post-AGB phases (Table 2, Iben 1984),
obtaining different guesses for FF occurrence in 10 to 21 $\%$ of all stars 
in this mass range, 
and up to about 60 $\%$ when he includes stars that {\it leave the AGB burning helium}.
Our parametrization is comparable with Iben's prediction. In fact, if we
were to chose that 60 percent of all stars (in this mass range) have the 
chance to experience a final flash, then about 12 $\%$ of the PNNi are in 
He-burning phase as observed in a synthetic HR diagram.

\subsection{The duration of the planetary nebula phase}

Following common wisdom, the duration of the PN phase is $\sim 30,000$
yr (e.g., Phillips 1989, and references therein).
This is derived from the size of the largest observed PNs ($\sim 0.7$
pc) coupled with a typical nebular expansion velocity of $\sim$25 km
s$^{-1}$. In
practice this assumes that all PNs remain visible as such until the
expanding nebula has reached the maximum observed size, or,
equivalently, that all PNs evolve in nearly the same way.
On the other hand, the mere fact that PNs
are produced by precursor stars in such a very extended range of initial
masses
($0.85\lesssim\mi\lesssim 9\,\smun$) makes most unlikely that
all
PNs have the same lifetime. Simple arguments suggest that the time
$t_{\rm max}$ after the cessation of the superwind during which a PN
remains observable 
scales as
$t_{\rm max}\propto M_{\rm PN}^{2/5}SB_{\rm min}^{-1/5}$, where 
$M_{\rm PN}$ is the nebular mass and $SB_{\rm min}$ is the minimum
surface brightness for a PN to be detected (P2,P5).
In the adopted mass loss parameterization, $M_{\rm PN}$ corresponds to
the
mass ejected during the AGB superwind phase, and ranges from $\sim
0.02\,\smun$
for $\mi=0.85\,\smun$, to over $1\,\smun$ for $\mi=8\,\smun$. This
factor of $\sim 50$ in PN mass therefore translates into at least a
factor $\sim 5$ in
$t_{\rm max}$. To get the actual duration of the PN phase we must
subtract to $t_{\rm max}$ the AGB to PN transition time, during
which the object is a proto-PN, i.e., $\tpn=t_{\rm max}
-\ttr$. If
$\ttr>t_{\rm max}$, then when the central star has reached 10,000 K
the nebular material has already dispersed, no observable PN is
produced, and one has a so-called {\it lazy} post-AGB remnant (P2).
To explore the effects of this subtle interplay between central star
and (partially decoupled) nebular evolution, we have explored the case of 
mass dependent $t_{\rm max}$, scaled with  $M_{\rm PN}$. The nebular mass was
derived by imposing its dependence on the plateau luminosity, since it is
at the reach of a critical luminosity that the shell is ejected. The parametrization of \mpn and $t_{\rm max}$ are described in the Appendix
(Eqs. A1 and A2). Simulations of post-AGB populations with constant maximum
PN ages 
will be also shown in this paper.

\subsection{The Montecarlo procedure}

Our population synthesis code starts with the option for a time limited
or luminosity limited sample. The former option is 
to explore post-AGB populations within a fixed time range,
(we are talking here about the stellar evolutionary time, 
not the duration of the PN phase, described in $\S$2.6).
This option is used to deduce the mass distribution of the PNN 
as derived from the synthetic population analysis.  
The latter option, the luminosity-limited sample, is used 
to compare directly synthetic and observed diagrams. 
All stars in the synthetic population would have L$ \gtrsim \lf$ (we assume 
$\lf$=1.0 \lsun)\footnote{Naturally, the luminosity limit can be varied 
accordingly to the observed situation we may want to reproduce.}.

To build the synthetic population, we proceed as follows:
 
\subsubsection{Time limited sample}

1) First, a value of the initial mass $\mi$
is extracted, following the distribution:

$$\psi(\mi)\propto\mi^{-(1+x)}\eqno(13)$$
 
2) The extracted
value of $\mi$ is entered into the initial mass-final mass relation to
get the mass $M$ of the post-AGB object. 

3) To proceed, we need to evaluate \mer~ (see $\S 2.4$), 
and correspondingly the wind
time scale can be calculated (see Eq. 9).

4) A random time is then extracted, within a chosen range ($0\lesssim t \lesssim
t_{\rm lim}$).

5) If the extracted time $t$ is 
larger than the wind timescale, we put $t=t-t_{\rm w}$,
and this value of $t$ is entered into the routines of $\S$2.1 
thus getting the corresponding location in the HR diagram. 
If $t$ is smaller than $t_{\rm w}$ the star is still in the wind phase, 
and we assign \teff=3.7 and $L=\lpl$. We call the resulting simulated star
a {\it wind} object\footnote{
Given our assumption on the MLR, 
we distinguish between wind objects and proto-PNN (in [9], below):
wind objects are those transition objects whose evolution is
wind dominated, proto-PNNi are all the other transition objects,
located beyond the AGB. The assigned temperature for the
wind object is rather arbitrary, but homogeneous to our mass loss choices.
Even if we were to locate the wind objects within the transition area, the
results would be not majorly affected, given the paucity of the samples. 
In a future paper, different mass loss choices will analyze the possible differences. Observationally, it may be hard to distinguish between these two
types of objects.}.

If we do not consider a FF, we go to point (8).
 
6) Should we consider the effects of a final helium flash,
another value of time $t_{\rm FF}$ is extracted, with 
$0 \lesssim t_{\rm FF} \lesssim\ t_{\rm lim}$, 
so as to get a random value for the time at which the flash takes place
\footnote
{Note that Iben (1984) has calculated slightly different probabilities for FF 
occurrence at different phases, but we use a flat probability for simplicity.}.

7) If $t<t_{\rm FF}$, then this $t$ is entered into the routines of
$\S 2.1$ to get the luminosity and temperature of the star
(in this case we have extracted a post-AGB star which
{\it will} experience a final flash, but the flash has not taken place
yet). 

If $t>t_{\rm FF}$, then the time elapsed after the  
the flash ($t'=t-t_{\rm FF}$) is entered into Equation (12)
to get luminosity of the post-FF star. To calculate the temperature we use 
the routines of $\S2.1$. 

8) The transition time $\ttr$ is obtained following the prescriptions 
in \S 2.4.

9) The object is finally classified into one of the following 
classes,
and plotted with a different symbol for each class. The object is 
classified as a proto-Planetary Nebula Nucleus (proto-PNN) 
if $t\le \ttr$. It is classified as
a PNN if $t+\ttr\le t_{\rm max}$, or $t'+t_{\rm FF}+\ttr\le
t_{\rm max}$, with the further distinction between objects that have
experienced a final flash (PNNHe) and those which did not (PNNH). It
is classified as a post-PNN object otherwise, again distinguishing
post FF and no post FF objects.

\subsubsection{Luminosity limited sample}

To build a luminosity-limited sample, we follow a similar procedure as in the
time-limited sample, except in (4), the extraction of a random time, 
that is chosen in the range  $0\lesssim t\lesssim \tf(M)+t_{\rm w}$,
where $\tf$ is the fading time to the minimum luminosity, $L=1.0~\lsun$
(see A3). 

The we proceed as in points (5) through (8) of the procedure in $\S$ 2.7.1, 
with the difference that  
$t_{\rm FF}$ is extracted within the new time limits.

\section{Results}

Our code is flexible, suitable to produce post-AGB population synthesis for 
many applications and many sets of parameters. Far from being exhaustive, this
paper includes only a small part of the possible applications. We have explored 
the issues and questions that we though to be of wide interests, and allow
to advance in out post-AGB evolution knowledge. Other applications will be 
developed in the future.

\subsection{Synthetic diagrams and mass histograms: 
the {\it basic} model}

The first synthetic population shown here is a 
luminosity-limited sample
of post-AGB stars. We extract 1500 objects, and we separate the 
{\it wind} objects, the PNNi, the 
proto-PNNi, and the post-PNNi following the prescriptions of Chapter 2.
In Figure 15 we show the synthetic HR diagram, where 313 objects
of the 1500 extracted are in the PNN phase, 21 are in the wind
phase, and 9 are proto-PNNi. The remaining objects
are, following our prescription, already in the post-PNN phase.

In the simulation of Figure 15 the residual envelope mass, \mer, is chosen to
be function of the core mass, as
described in $\S$2.4; we have
used Weidemann's IMFMR, Salpeter's 
IMF, and no FF objects (i.e., all PNN are hydrogen
burners). 
The maximum PN time has been set following Eq. A2, with K=4$\times 10^5$,
a relatively long time for PNe to disappear. We use this time through $\S$3.5, considering furtherly the effects of the maximum
PN age in $\S$3.6.
We define the simulation plotted in Figure 15 as {\it basic}, 
and we 
will explore the parameter space by changing one parameter at the time 
with respect to this basic population.

As expected, the synthetic stars cluster toward the lower masses,
only the very low luminosity part being populated by higher masses.
This is exactly the effect we see in all {\it complete} galactic 
PNNi samples.

For more insight, let us plot the corresponding distribution of the 
visual magnitudes versus evolutionary timescale (Fig. 16). 
We translate temperatures and luminosities
into visual magnitudes by following the bolometric correction from 
Code \etal (1976). As already evident in the previous plot,
there are two mainly populated loci of the \mvt~plane as well, corresponding one to 
{\it plateau} luminosity of the low-mass objects, 
and the other to the stellar
crowding toward low luminosities of the intermediate- and high-mass objects. 
The large fraction of PNNi that have decayed to 
post-PNNi at late evolutionary times is very evident in this Figure. 
This effect is mass-dependent, and smaller mass PNN {\it disappear} earlier in their life.  

In Figure 17 we show the (logarithmic) mass distribution for the {\it basic} 
synthetic population.
This distribution has been obtained by
running a sample of 1500 post-AGB stars with maximum HR diagram-life time of 30,000 years. The logarithmic scale has been chosen for a better view of the
distribution in the whole mass range.
The mass distribution of Figure 17 shows the clear clustering of model
post-AGB stars around $\sim$0.6 \sm, and a gradual spread to higher masses. 

Due to the particular choice of the time interval explored ($t<30,000 
~yr$), we do not obtain post-PNNi with this time-limited selection.
In Table 2 we show the composition of a selection of the
synthetic populations shown in this paper. The columns list, respectively, 
the number of PNNi, wind objects, post-PNNi, and proto-PNNi 
stars for each extraction. Composition of typical populations may vary 
from one random extraction to another. From Table 2 we can directly compare the time-limited to the luminosity-limited samples.

\subsection {Effects of the initial mass function}

To test the variations of post-AGB distribution for 
different Initial Mass Function we have run synthetic populations with
the following IMFs (see $\S 2.2 $ for meaning of variable x and for SF
choice): 
(1) the standard, Salpeter's (1955) IMF, 
with $(1+x)=2.35$ in the whole mass range considered; 
(2) Miller \& Scalo's (1979) IMF, 
with $(1+x)=1.4$ for $M<1$\sm~ and $(1+x)=2.5$ for $M>1$ \sm;
(3) the IMF by Kroupa \etal (1991),
with $(1+x)=1.85$ for
$M<1$ \sm~ (this distribution is not defined for larger masses), 
and (4) an updated empirical parametrization by
Scalo (1998) with $(1+x)=1.2$ for $M<1$\sm~ and $(1+x)=2.7$ for $M>1$\sm. 

In Figure 18 we show HR distributions accounting for IMF variations, when the IMF is the only parameters 
changing across the panels. In each case, we have extracted a
luminosity-limited sample of 1500 
post-AGB stars, with Weidemann's IMFMR, and mass-dependent 
residual envelope mass. The stars do not experience a final helium shell 
flash. In the case of Kroupa \etal's low mass IMF, we use Scalo's (1998) IMF
for masses larger than solar. 

By examining Figure 18 \footnote{In Fig. 18, and after, we use the
following conventional nomenclature for figure panels: 
a=lower left, b=upper left, c=lower right, d=upper right. Panel (a) always
shows the {\it basic} sample.} we see very little changes on the \logl-
\logt plane for different IMF in the mass range considered.
The PNN populations in the case of variable IMF index (Fig. 18bcd) 
are larger for low 
masses respect to the {\it basic} model. 
The proto-PNN population changes slightly (we are dealing with low number 
statistics, anyway), but overall the distributions
are similar in all four cases of Figure 18.
Better to say, that the HR diagram is not the ideal way to pick up these differences, occurring in this case at low-luminosities, in a crowded
part of the diagram.
Observations of real situations such as those of the four panels of Figure 18 would not be distinguishable from one another.

In Figure 19 we illustrate the mass distribution
for the synthetic populations corresponding to Figure 18, each panel 
showing a different IMF. Naturally, time-limited 
samples have been used.
It is worth noting that peaks in the distribution of less than $\sigma$
can be produced by the randomness of the simulation. 

The effects of the IMF are noticeable given the large sample of model
stars. For example, Scalo's IMF (Fig. 19d) would produce a broader distribution
toward the higher masses. Note that the logarithmic scale partially hides
the fact that all these mass distributions are extremely peaked 
around 0.55 \sm. 

\subsection {Effects of the initial mass-final mass relation}

As introduced in $\S 2.3$, we will use four different IMFMR and test their effects on the post-AGB populations. In Figures 20 and 21 we plot the results relative to (a) Weidemann's, (b) Ciotti \etal,
(c) Herwig's, and (d) IR's IMFMRs. Figure 20 shows the 
distributions on the HR diagram. On this plane, the sample with IR's IMFMR stands aside, since the IR's prescription produces more massive post-AGB. 
 
The differences among the other distributions are more evident in Figure 21,
where we plot the visual magnitude versus evolutionary time of the 
synthetic stars. 
By comparing the {\it basic} and Herwig's IMFMR diagrams, above Mv=5 the two distributions are similar, but the differences 
appear and get more evident for fainter objects. 

To compare these results with the actual observed stars is beyond the purpose 
of this paper. Nonetheless, we can state that the different IMFMRs could 
be inferred mostly at low brightness, with consequent higher uncertainty
of the comparison with the data. 

The mass distributions of Figures 22 set aside once again the IR
IMFMR. This old parametrization allows a continuum of masses between 0.55 
and 1.4 \sm. The other 3 cases have similar low-mass distributions,
while the high-mass distributions are similar in the Ciotti \etal's
and the Herwig's IMFMRs, as the {\it basic} sample allows for few very high mass stars. The observational consequences of the four different IMFMRs are
mainly constrained in the high-mass tail of the distributions. 

\subsection{The residual envelope mass}

The envelope mass that is left on the star after the envelope ejection,
\mer, plays a fundamental role in the following stellar fate.
Our {\it basic} model uses \mer=f(M), with an ad hoc
parametrization (see $\S 2.4$).
Since the value of \mer~ is undetermined, as well as its distribution
with respect to the other physical parameters, we chose to explore scenarios
of post-AGB evolution both with 
with variable and constant \mer. We also
produce populations with 
random residual envelope masses, to represent the extreme (but not unrealistic)
indetermination of \mer,
and consequently of the transition time.

In Figure 23 we show the results with the different scenarios:
the {\it basic}
model (a) 
has mass-dependent \mer, model (b)  
has random \mer~ (with \mer$<$0.1 \sm), 
model (c) also has random \mer~ (with \mer$<1 \times 10^{-4}$ \sm),
and model (d) has constant \mer=$5 \times 10^{-3}$ 
\sm, independent on the progenitor mass (lazy-AGB stars have not been
produced in these simulations, given the choice of K in equation A2).

The effects of the indetermination of the residual envelope mass are
striking on the resulting post-AGB populations.
A complete description on mass loss
for low- and intermediate-mass stars does not exist to date.
Our experiments (e.g., Fig. 23)
represent a way to determine the effects of the
mass dependence on the MLR on observable sets of PNNi and related objects.
Once again, in this paper we do not compare these effects with the
observed stars, we just set the stage for future comparisons.
When using \mer=5$\times 10^{-3}$\sm, the distribution is very similar to
the basic case. 

We use the random \mer~ to further analyze the effects of the mass loss
on observed populations. 
The distributions that we obtain as a result of the simulations with 
random \mer~ are very different depending on the upper limits that we 
set for \mer. For instance, for \mer$<1 \times 10^{-4}$ \sm, no wind
nor proto-PN are created. Furthermore, very few high-luminosity PNe are 
available with this low-limit random \mer~ (see also Table 2).
These effects
could be observed in PNNi and proto-PNNi at known distances.
At lower luminosities, the two distributions look similar 
especially if we take into account observing errors.

The mass distributions of post-AGB populations with different choices of \mer~ are
shown in Figure 24. The effects of changing \mer~ are even more extreme
in these time-limited samples. Let us examine Figures 24b and 24c (upper-left
and lower-right panels). In both panels, \mer~ has random values, 
the difference being the maximum allowed \mer. 
In the simulation of 24b, where the 
maximum residual envelope mass is 0.1 \sm, the distributions
is dominated by the objects in the wind phase, and the lack of PNNi (actually,
the simulation produces two PNNi, hiddent in the log-representation).
The most notable
characteristics of 
panel 24c are the lack of wind objects or proto-PNNi. The very low \mer~ make the transition time very short,
and, as a consequence, the lack of proto-PNNi. Studies of a sample
including OH/IR
and other transition objects, and PNNi within the same environment, should reflect the mean value of \mer~ in the sample.

In Figure 24d most post-AGB stars
belong to the pre-Planetary Nebula 
(wind and proto-PNN) phases. Only a few
percent of the stars are PNNi. The situation of Figure 24d has been 
produced only to
show an extreme case, since it is highly unrealistic that all stars have the same
residual envelope mass, independent of their initial mass on the main sequence,
and on their core mass.

\subsection{The final helium-shell flash}

In this paper we did not use the helium-burning (nor a combination
of hydrogen and helium burning) stellar models 
to determine the effect of a helium-burning PNN population. The parametrization that we use is 
described in $\S 2.5$, and it agrees with observations. Following, we show how
the presence of helium-burning stars can effect the
observable post-AGB populations.
In Figure 25 we show the synthetic population on the HR diagram for a sample 
of 1500 post-AGB stars in which
20 $\%$ of the post-AGB stars have experienced a final flash. The fraction
of stars in post-FF is chosen in agreement with the guesses based
on the evolutionary models by Iben (1984). 
We compare this population with the {\it basic} model.
We note, from Table 2,
that the overall composition of the sample in terms of ratio of PNNi 
to other components is (statistically) very similar to 
the {\it basic} sample.

In Figure 25 we have separated the He- and H-burning stars. Panel (a) shows 
the {\it basic} model (1500
objects with Salpeter's IMF, Weidemann's IMFMR, \mer=f(M), and no
post-FF). In the other panels, we show (b) the 
post-AGB stars having experienced a FF (20$\%$ of the sample);
(c) stars 
that are in H-burning, post H-burning and/or {\it will} experience
FF (the latter group of stars is observationally indistinguishable from H-burning stars); (d) the composite 
population of panels (b) and (c), thus the observable post-AGB population.
Figure 26 shows the same populations of Figure 25 in the \mvt plane.
The simulated He-burning populations stand out for lack of low-brightness PNNi.

It is not easy to observationally discern among 
H- and He-burning stars. The problem is that the stellar abundances are not
easy to measure in PNNi. In general, H-depleted post-AGB stars are believed to
be He-burners, but it is not so easy to single out the H-burning stars,
as H-rich post-AGB could indeed be He-burners as well as H-burners. 
The simulation of Figures 25 and
26 can help us in this respect. In fact, the ratio of PNNi brighter than, say,
5 magnitudes in the basic sample is about a third of the total sample, while 
in the composite population (Fig. 26c) this ratio goes up to half the sample.
A homogeneous, complete observed PNN sample should show this kind of 
discrepancy.

\subsection{The planetary nebula life time}

Through this paper we have assumed that the maximum age for a PN 
depends on the PN mass, as in Equation A2, with constant K=4$\times
10^5$. We have also produced other synthetic 
populations by keeping the nebular mass dependence, and changing the constant. If we keep
all other parameters as in the basic model, we obtain that the number of 
PNNi goes to zero as K is lowered to about 5$\times10^3$. This happens because
the shorter lifetime of PNe makes most of the post-AGB population be in a 
post-PNNi phase. In this case, a large number of lazy AGB stars are also produced. In general, lazy AGB stars are produced by running the
basic model with K$< 8 \times10^4$.

If we were to eliminate the nebular mass dependence from $t_{\rm max}$ the
situation changes noticeably.
In Figure 27 we show a simulated population with all parameters identical to
the {\it basic} model, except we have fixed the PN life time to 30,000 years
for all nebulae. The result is remarkably different than the basic
model (see Fig. 15 and Table 2). Most low-mass post-AGB stars are in 
post-PNN phase. If we were to lower the maximum PN life time to 10,000 years 
(not an 
unreasonable choice) we obtain that most PNNi in the diagram would
have a high mass.
This is in contrast with the 
observations, thus our simulations lead us to believe 
that the lifetime of PNe depends on the nebular mass. Detailed comparison
with homogeneous dataset should be used to confirm this inclination.

\section {Summary and future work}

We have used up-to-date evolutionary tracks as templates to build a code
for post-AGB population synthesis. Our models aim at understanding the 
{\it fine tuning} of post-AGB evolution, including the consequences of
IMF and IMFMR, the transition time and its correlation with
the residual envelope mass, the actual duration of the PN phase, 
the relative population of proto-PNNi, wind objects, and post-PNNi stars, and
the occurrence and timing of the FF phase. 
Our synthetic tracks, available to obtain \logt 
and \logl for post-AGB stars of any stellar
mass in the range $0.85 \lesssim\mi< 9\smun$, are obtained with very high 
precision in reproducing the actual evolutionary tracks. The interpolation
method is simple and fully explained in this paper, so that the readers
can reproduce the synthetic populations included here. 

In this paper we have shown a sample of the possible applications, 
without going into detailed comparisons with the observed data. 
Among the results shown here, we found that (1)
the dependence of
the synthetic populations on the assumed IMF and IMFMR is mild; the post-AGB
populations are not ideal indicators of the IMF. (2) The residual envelope mass, after the envelope ejection, has a 
strong effect in determining the subsequent post-AGB evolution; its 
indetermination
produces very high indetermination on the transition time, and
ultimately on the resulting post-AGB populations.
(3) The ratio of He- to H-burning
PNNi can be reproduced with population synthesis.

The central importance of this paper consists in showing that the many 
fundamental variables of stellar evolution have a major role in determining
post-AGB populations. The variation and indetermination of these parameters
should not be overlooked in comparing data and theory.

The theoretical work contained in this paper 
was implemented with the goal of being versatile and useful for
a full host of applications. Among other possible uses of these synthetic
tracks are the studies of evolutionary effects on the
Planetary Nebula Luminosity Function (PNLF), and how those translate in the 
variation of the extragalactic distance scale, as derived from the PNLF.
Stanghellini (1995) has shown in a preparatory work that the PNLF is at
variation with respect to the transition time. The updated models 
presented here allow the most detailed study of the PNLF on the evolutionary
parameters. 

Future applications also include the simulations of bulge, elliptical galaxy,
and Magellanic Cloud
populations, where a different treatment of the SF/IMF should be used, and
possibly different stellar chemistry. 

\acknowledgements

Thanks to Laura Greggio for her help in the interpolation techniques, 
to Antonella Nota and many others for their bibliographic indications, 
and to an anonymous referee for pointing out an error in the previous draft
and for many suggestions. 

\clearpage

\appendix

\section{Appendix}

In order to parametrize the duration of the PN phase, we 
assume that the nebular mass is 0.02 \sm~
for M=0.55 \sm~ (core mass), and  1.5 \sm~ for M=1.2 \sm. The 
planetary nebula mass can be described by 
the following correlation between nebular mass and {\it plateau} luminosity:

$$\mpn=4.495\times10^{-8} \lpl^{1.636}, \eqno(A1)$$

and

$$t_{\rm max} = K \mpn^{2/5}. \eqno(A2)$$

To calculate the fading time to $L=1.0 $\lsun, $\tf$,
 we use the VW tracks and we interpolate versus the (post-AGB) mass, obtaining

$$ \tf(M)=7.0957-2.1441 M. \eqno(A3) $$

\clearpage

\figcaption[]{Effective temperatures versus evolutionary time for M=0.569, 
0.597, 0.633, 0.677, 0.754, and 0.9 \sm.\label{fig1}}

\figcaption[]{Lumonisities versus evolutionary time for M=0.569, 0.597,
 0.633, 0.677, 0.754, and 0.9 \sm.\label{fig2}}

\figcaption[]{Effective temperature versus normalized age
in the hydrogen burning phase.
\label{fig3}}

\figcaption[]{Effective temperature versus normalized age
 in the quenching phase.
\label{fig4}}

\figcaption[]{Normalized effective temperature versus normalized age
in the cooling phase.
\label{fig5}}

\figcaption[]{Normalized effective temperature versus normalized age
in the white dwarf phase.
\label{fig6}}

\figcaption[]{Luminosity versus normalized age 
in the Hydrogen burning phase (or plateau).
\label{fig7}}

\figcaption[]{Normalized luminosity versus normalized age
in the cooling phase, for M=0.569,
0.597, 0.644, and 0.644 \sm.\label{fig8}}

\figcaption[]{Luminosity  versus normalized age
in the white dwarf phase.\label{fig9}}

\figcaption[]{The IMFMR: from IR (solid line),
Ciotti \etal (1991, long dashed line),
Weidemann (1987, short-dashed line), and Herwig (1997, dash-dotted line).
\label{fig10}}

\figcaption[]{The transition time (thick solid line), obtained 
with \mer~ as a function of 
the core mass, as explained in $\S$3.1. Also plotted are the wind (thin solid line), and nuclear (dashed line) time scales. 
\label{fig11}}

\figcaption[]{As in Figure 11, but with \mer=3$\times 10^{-4}$
 \sm.\label{fig12}}

\figcaption[]{As in Figure 11, but with \mer=5$\times 10^{-3}$  \sm.\label{fig13}}

\figcaption[]{As in Figure 11, but with random \mer~ $<5e-3$ \sm.\label{fig14}}

\figcaption[]{The {\it basic} post-AGB synthesis, luminosity-limited.
Solid circles = PNNi; squares = {\it
wind} object; open circles = proto-PNNi. 
The tiny dots show the locations of the post-PNNi.
Tracks correspond to M=0.535, 0.569, 0.597, 0.633, 0.677, 0.754, and 0.9 \sm.
\label{fig15}}

\figcaption[]{The {\it basic} post-AGB synthesis (luminosity-limited distribution),
on the \mvt~ diagram. Solid circles = PNNi; squares = {\it
wind} object; open circles = proto-PNNi. 
The tiny dots show the locations of the post-PNNi.
Tracks correspond to M=0.535, 0.569 and 0.9 \sm.\label{fig16}}

\figcaption[]{Logarithmic mass distribution for the {\it basic} sample
(time-limited distribution). 
Solid line = PNNi;
broken line = proto-PNNi; shaded histogram = wind objects.\label{fig17}}

\figcaption[]{Synthetic post-AGB populations on the HR diagram. The IMFs are from
(a) Salpeter (1955) (b) Miller \& Scalo (1979); (c) Kroupa \etal
(1991); and (d) Scalo (1998). 
Other parameters are as in the {\it basic} model, which is panel (a).
Symbols and tracks as in Fig. 15.\label{fig18}}

\figcaption[]{Mass distribution of synthetic populations with different IMFs. 
The IMFs are from
(a) Salpeter (1955); (b) Miller \& Scalo (1979); (c) Kroupa \etal
(1991); and (d) Scalo (1998). 
Other parameters are as in the {\it basic} model.
Histogram lines as in Fig. 17.\label{fig19}}

\figcaption[]{The \logt-\logl plane population synthesis for (a) Weidemann's, 
(b) Ciotti \etal's, (c) Herwig's, and
(d) IR's IMFMR. All other parameters are as in the {\it basic}
models.
Symbols and tracks as in Fig. 15.\label{fig20}}

\figcaption[]{The \mvt~ population synthesis for (a) Weidemann's (1987), 
(b) Ciotti \etal's (1991), (c) Herwig's (1997), and (d) 
Renzini \& Voli's (1981) IMFMR.
All other parameters are as in the {\it basic} model.
Symbols and tracks as in Fig. 16.\label{fig21}}

\figcaption[]{The mass distributions of populations with different IMFMR:
(a) Weideman's, (b) Ciotti \etal's, (c) Herwig's, and (d) 
Renzini \& Voli's IMFMR.
All other parameters are as in the {\it basic} model.
Histogram lines as in Fig. 17.\label{fig22}}

\figcaption[]{The \mvt~ relation for different choices of the \mer; (a):
\mer=f(M); (b): random \mer, with \mer$<$0.1 \sm; 
(c): random \mer, with \mer$<1\times 10^{-4}$ \sm; 
(d): constant \mer=5$\times 10^{-3}$ \sm.
All other parameters are as in the {\it basic} model.
Symbols and tracks as in Fig. 16.\label{fig23}}

\figcaption[]{Mass distribution for the choices of \mer~ as in 
Fig. 23:
(a):
\mer=f(M); (b): random \mer, with \mer$<$0.1 \sm; 
(c): random \mer, with \mer$<1\times 10^{-4}$ \sm; 
(d): constant \mer=5$\times 10^{-3}$ \sm.
All other parameters are as in the {\it basic} model.
Histogram lines as in Fig. 17.
\label{fig24}}

\figcaption[]{The effects of a final helium-shell flash during cooling, 
on the HR diagram. 
(a) {\it basic} population; (b) post-FF stars; 
(c) H-burning or post H-burning stars; 
(d) population of post-AGB stars with 20$\%$ objects in post-FF phase
(i.e., b+c populations). 
All other parameters are as in the {\it basic} model.
Symbols and tracks as in Fig. 15.\label{fig25}}

\figcaption[]{The effects of a final helium-shell flash during cooling, 
in the \mvt~ diagram. 
(a) {\it basic} population; (b) post-FF stars; 
(c) H-burning or post H-burning stars; 
(d) population of post-AGB stars with 20$\%$ objects in post-FF phase
(i.e., b+c populations). 
All other parameters are as in the {\it basic} model.
Symbols and tracks as in Fig. 16.\label{fig26}}

\figcaption[]{The post-AGB synthesis, luminosity-limited, with PN mass
independent maximum PN lifetime ($t_{\rm max}$=30,000 yr).
Symbols and tracks as in Fig. 15. 
\label{fig27}}
\clearpage

\begin{table}
\caption[]{Characteristics of evolutionary tracks phases}
\begin{tabular}{lrrrr}\hline \\
Phase&  $M$&  $t_{\rm end}$& ${\rm log}~T_{\rm eff, end}$&
${\rm log}~L_{\rm end}$  \\

& [\sm] &   [yr] &  [K] & [\lsun] \\
\hline

{\it H-burning}&  0.569& 3.215$\times 10^4$&  4.907& 1.47\\
& 0.597& 1.098$\times 10^4$& 5.153 & 3.19 \\
& 0.633 &  4448 & 5.238 & 3.348 \\
& 0.677 &  2667 & 5.292 & 3.474 \\
& 0.754 &  1067 & 5.39 & 3.699 \\
& 0.9 &  141.7 & 5.581 & 4.02 \\
{\it quenching}& 0.569 & 3.614$\times 10^4$& 5.038 & 2.579 \\
& 0.597 & 1.214$\times 10^4$& 5.071 & 2.458 \\
& 0.633 & 4862 & 5.112 & 2.424 \\
& 0.677 & 2925 & 5.162 & 2.527 \\
& 0.754 & 1177 & 5.315 & 3.033 \\
& 0.9 & 167.1 & 5.403 & 2.953 \\
{\it cooling}& 0.569 & 3.031$\times 10^5$& 4.907 & 1.47 \\
& 0.597 & 3.258$\times 10^5$& 4.9 & 1.297 \\
& 0.633 & 3.113$\times 10^5$& 4.906 & 1.203 \\
& 0.677 & 3.165$\times 10^5$& 4.911 & 1.125 \\
& 0.754 &  3.318$\times 10^5$& 4.91 & 0.919 \\
& 0.9 &  2.21$\times 10^5$& 4.952 & 0.855 \\
{\it white dwarf}& 0.569 & 3.068$\times 10^6$& 4.659 & 0.112 \\
& 0.597 & 4.044$\times 10^6$& 4.62 & -0.127 \\
& 0.633 & 4.681$\times 10^6$& 4.596 & -0.299 \\
& 0.677 &4.973$\times 10^6$& 4.583 & -0.411 \\
& 0.754 & 7.012$\times 10^6$& 4.547 & -0.683 \\
& 0.9 & 1.164$\times 10^7$& 4.545 & -0.877 \\
\hline\\
\hline
\end{tabular}
\end{table}

\clearpage
\begin{table*}
\caption[]{Composition of typical synthetic post-AGB populations}
\begin{tabular}{lrrrr}\hline \\
 run& PNNi &  wind objs.& post-PNNi& proto-PNNi  \\
\\
\hline\\
Luminosity-limited sample ($L>1\lsun$) \\

{\it basic} & 313 & 21 & 1157 & 9 \\
Miller \& Scalo	&	246&	18&	1230&	6\\
Kroupa \etal (1991)	&	265&	13&	1212&	10\\
Scalo (1998)	&	282&	14&	1199&	5\\
Ciotti \etal (1991)	&	401&	10&	1083&	6\\
Herwig (1995)	&	352&	12&	1125&	11\\
Renzini \& Voli&        585&  15&  895&   5\\

random \mer ($<$ 0.1 \sm)	 &	191&	520&	784&	5\\
random \mer ($< 1\times 10^{-4}$ \sm)	&	226&	-&	1274&	0\\
\mer= $5 \times 10^{-3}$ \sm&	232& 41& 1225& 2\\

20 $\%$ post-FF  & 275& 11& 1210& 4\\
$t_{\rm max}$=30,000 yr& 52& 17& 1426& 5\\
\hline\\

Time-limited sample  ($t_{\it lim}<30,000$ yr) \\

{\it basic}	&		940&	397&	-	&163\\
Miller \& Scalo &		959&	400&	-	&141\\
Ciotti \etal (1991)&		975&	381&	-	&144\\
Herwig (1995)&     970&  288&  -& 142\\
random \mer ($<$ 0.1 \sm)&	2&	1498&	-	&-	\\
random \mer ($< 1\times 10^{-4}$ \sm)& 1215& -& 285& -\\
\mer=$5\times 10^{-3}$ \sm& 578& 813& 8& 101\\

\hline\\
\end{tabular}
\end{table*}


\end{document}